\documentstyle[twocolumn,aps,epsf]{revtex}
\begin{document}
\newcommand{\beq}{\begin{equation}}
\newcommand{\eeq}{\end{equation}}

\title{Coarse grained description of the protein  folding}

\author{Marek Cieplak and Trinh Xuan Hoang}

\address{Institute of Physics, Polish Academy of Sciences,
02-668 Warsaw, Poland}

\address{
\centering{
\medskip\em
{}~\\
\begin{minipage}{14cm}
We consider two- and three-dimensional lattice models of proteins which
were characterized previously.  We coarse grain their folding dynamics by
reducing it to transitions between effective states. We consider two
methods of selection of the effective states. The first method is based on
the steepest descent mapping of states to underlying local energy minima
and the other involves an additional projection to maximally compact
conformations. Both methods generate connectivity patterns that allow to
distinguish between the good and bad folders. Connectivity graphs
corresponding to the folding funnel have few loops and are thus tree-like.
The Arrhenius law for the median folding time of a 16-monomer sequence is
established and the corresponding barrier is related to easily identifiable
kinetic trap states.
{}~\\
{}~
\end{minipage}
}}

\maketitle

Proteins that are found in nature are special sequences of aminoacids that
fold rapidly into their native states under physiological
conditions\cite{1}.  The native states control functionality of proteins
and are commonly assumed to coincide with the ground state conformations.
Exploration of the protein's phase space in search of the native state
typically  takes milliseconds.  This is in contrast to an essentially
indefinite search expected for randomly  constructed sequences of
aminoacids -- such sequences are generally bad folders.  It is believed
that the well folding biological sequences have an  energy landscape  with
a dominating folding funnel which restricts  the number of visited
conformations during folding.   Based on simulations of lattice models,
Onuchic et al.\cite{2} have identified the crucial  landmarks in the
folding funnel such as the molten globule states and low energy
bottlenecks. Before, Leopold et al.\cite{3} have  studied  transition rates
between conformations found in the last stages of folding and interpreted
the resulting trajectories as forming a folding funnel.  In this paper, we
focus on ways to define folding funnels operationally, in numerical
simulations. We consider two- and three-dimensional lattice models,  the
dynamics of which are given as a Monte Carlo process with single and two
monomer moves.  

The point of view that we propose here is that the Monte Carlo dynamics
generates too many states, each of an overall negligible  occupancy, to
allow for a  convincing and easy to do identification of the funnel without
some educated organization of the data.  Thus some reduction in the
description should facilitate the task by an elimination of conformations
that are less relevant. A valid analogy here is with a solid: understanding
properties of a solid  can often be reached from knowledge of the crystal
structure in the ground state without taking into account any phonon
states.  Stillinger and Weber \cite{4}, in the context of  glasses, and
Cieplak and Jaeckle\cite{5}, in the  context  of spin glasses,   have
accomplished  a useful elimination of such 'phonon' states  by mapping
states of  a system  to  underlying local energy  minima obtained, in a
unique way, through the steepest descent method. The motion of the system
through the phase space could then be viewed as an effective migration
between the underlying 'valleys'.  This approach has been subsequently
adopted to proteins by Cieplak, Vishveshwara, and Banavar\cite{6} and by
Cieplak and Banavar\cite{7}. It has been implemented for  two-dimensional
16-monomer  lattice models.  The procedure consisted  of a two-stage
mapping: first an encountered conformation was mapped to a  local energy
minimum  (LM) through the steepest descent method and then the LM was
mapped to a nearest maximally compact conformation, called 'cell' for
short, defined as one which has maximum energy overlap in common contacts
between LM and the cell. If several cells satisfy this criterion, the one
with the lowest total energy is picked. 

This particular scheme of  coarse graining  of the protein dynamics  has
been  successful since the resulting pattern of connectivities between
frequently occupied cells  was clearly differentiating between the bad and
good folders  and, in the latter case, was showing emergence of a funnel.
It has turned out, however, that the approach based on cells  is difficult
to implement for longer polymers, especially in three dimensions. For
instance, for a 27-monomer lattice model, there are 103346 maximally
compact $3\times 3 \times 3$ cells and it is hard to find a fast way  to
tell which of them is the closest to an LM.  An alternative to the
criterion of  maximum energy overlap is a mapping to a cell which is the
easiest to be reached kinetically but finding a reasonable algorithm for
this has turned out to be even harder.  More importantly, as we shall see
here, the connectivity patterns usually are not sensitive enough to allow
for a detection of truly relevant trapping states.

In this paper, we discuss a coarse graining scheme that is based  on the
LM's instead on the cells. Since the LM's are much more abundant than cells
it might seem that the technical problems compound.  What makes the scheme
tractable, however, is that we do not consider all LM's that the system is
endowed with, but only those which have been encountered. In addition, this
allows for a more detailed characterization of funnels.

We start, in Section 2,  by  considering  two 12-monomer sequences, A and
B, the dynamics of which have been recently studied in two dimensions by an
exact solution of the master equation\cite{8}.   The properties of these
sequences are then very well understood and, in particular, the kinetic
traps that govern the long time dynamical behavior at low temperatures,
$T$, have been identified. A is a good folder whereas B is a bad one and we
demonstrate that the pattern of connectivities between LM's into which the
states of the system have been mapped to yields a funnel for A but not for
B.   In Section 3, we consider two 16-monomer sequences in two dimensions,
R and DSKS' of references \cite{6,7} -- again the good and bad folder
respectively, and compare the LM-based dynamics to the cell-based dynamics.
Finally, in Section 4, we present results for  one good folder in 3
dimensions.

The energy of all of these sequencies is  given by
\begin{equation}
E\; = \; \sum _{i<j} \; B_{i,j} \; \Delta (i-j) \;\;\;,
\end{equation}
where $\Delta(i-j)$ means that the monomers $i$ and $j$ form a contact,
i.e. they are nearest neighbors on a lattice but are not neighbors along
the sequence.  $B_{i,j}$ are the corresponding contact energies --
essentially the numbers generated with the Gaussian probability
distribution but with a mean shifted to negative values to provide
compactness in the ground state. The equilibrium properties of the
sequences may be characterized by a folding temperature, $T_f$.
Following\cite{9} we define $T_f$ as $T$ at which the equilibrium
probability to find the system in the native state crosses $\frac{1}{2}$.
A large value of $T_f$ signifies substantial thermodynamic stability which
good folders are expected to possess. Temperature scales that characterize
dynamics can be determined from the plot of the median folding time,
$t_{fold}$ versus $T$: $T_{min} $ is where $t_{fold}$ achieves a minimum
and $T_g$, the glass temperature, is a low temperature point at which
$t_{fold}$  becomes steep. The definition of $T_g$ depends on adoption of a
cut-off  time that is considered to be too long  whereas $T_{min}$,
corresponding to the fastest folding, is criterion-independent. For good
folders $T_f$ is comparable to $T_{min}$. For bad folders $T_f$ is much
smaller than $T_{min}$ and then the system becomes trapped in a non-native
state at low temperatures before acquiring any substantial stability of the
native state.  The characterization of the two-dimensional sequences
studied here has been given before and we focus only on the coarse-grained
kinetics.

The Monte Carlo process starts from a random self-avoiding walk and it has
the move sets of Figure 1a  and case (i) of  Figure 1b in
reference\cite{10}. The single monomer moves have an a priori probability
of 0.2 and two monomer moves that of 0.8.  This process is performed in
such a way that the detailed balance condition is satisfied\cite{8,10},
i.e. besides the energetics, the probability to perform a move depends on
how many kinetic possibilities are allowed in a given conformation,
compared to the maximum number of possibilities of $N+2$, where $N$ is the
number of monomers in a sequence.

\section{{\bf  12-monomer sequences in two dimensions}}

The $N$=12 sequences A and B are defined in reference\cite{8}.  Both can
exist in 15037 conformations out of which 31 are $2\times 3$ cells.
Sequence A  has the native state, of energy -11.5031, which has an
appearance of the letter S.  Altogether, the sequence has 495 LM's out of
which 403 are V-shaped, i.e., any move out of them costs an energy
increase, and 92 U-shaped, i.e. some moves leave the energy unchanged.  All
states in an U-shaped minimum count as one  in what follows.   Sequence B
has a doubly degenerate native state of energy -8.7675 -- both states count
together when considering folding; none of the states in the doublet is
maximally compact.  Among the 496 LM's, 400 are V-shaped and 96 are
U-shaped.  $T_f$ and $T_{min}$ for sequence A are comparable since they are
around 0.7 and 1 respectively.   For the bad folder  B, on the other hand,
$T_{min} $ is again around 1 but $T_f$ is 0.01.   We compare dynamics of
the two sequences at two temperatures: 1 and 0.4.  For sequence A,
$t_{fold}$ at these two temperatures is 2 052 an 36 022 respectively and
for sequence B it is 2 457 and 215 364, as obtained by studying 500
different trajectories.

Figure 1 shows energy in a segment of  a  Monte Carlo trajectory for A at
T=1 and compares it to energies obtained by the one- and two -stage mapping
to the LM's and  cells respectively.  This temperature corresponds to the
fastest folding.  Naturally,  the higher level of coarse-graining, the
smoother the dependence of the effective energy on time.  For both methods
of coarse-graining,  the native state appears to have substantial occupancy
even though the sequence has not folded yet.   It is thus clear that the
system moves pretty much in the native valley which is easy to detect  if
one removes  the 'spurious' states from  the description.

Figure 2 shows $P_0$, $L_0$, and $C_0$ versus $T$ for sequence A.   The
first of these parameters is the equilibrium probability to find the native
state.   The second is the probability to find the native state after
mapping to the local energy minima. Finally, the third is the probability
to find the native state after mapping to the cell states.  We see that the
cell dynamics enhances the role of the native cell much more significantly
than the LM-based dynamics and the maximum $C_0$ for the good folder is
about 3 times as large as for the bad folder.  The maximum in $C_0$ is
closer to $T_f$ whereas the maximum in $L_0$ is closer to $T_{min}$.

All states of the system, local energy minima or not, can be enumerated and
their occupancies can be monitored.  Figure 3 shows occupancies of  states
found  during folding on  500 Monte Carlo trajectories for sequence A and
compares them to occupancies of LM's obtained by  the one-stage mapping.
At $T_{min}$, local energy minima  on the trajectories count overall as
much as other states.  In the low energy part of the spectrum, however, the
LM's dominate heavily. At low temperatures, on the other hand,  the time
spent in states which are not minima is negligible.   Furthermore, certain
LM's become heavily populated.  The biggest occupancy  belongs to the
state, denoted as TRAP,  which is the most potent kinetic trap on the way
to folding.  This is the same state which has been identified in
reference\cite{8} as contributing  most heavily to the eigenmode
corresponding to the longest relaxation time. Thus identification of the
kinetic traps does not need  to  involve diagonalization of the master
equation -- this task can be accomplished by counting  occupancies of
states encountered  in the Monte Carlo.  The trap state is also
substantially represented after mapping all states to LM's through the
steepest descent procedure.

When the trajectories are not terminated  at folding but are continued 
long enough (of order 1 million steps) to see equilibrium
values of $P_0$, the occupancy of the trap state drops from about 25\%,
at $T$=0.4, to about 3\% -- both before and after the mapping,

\begin{figure}
\epsfxsize=3.4in
\centerline{\epsffile{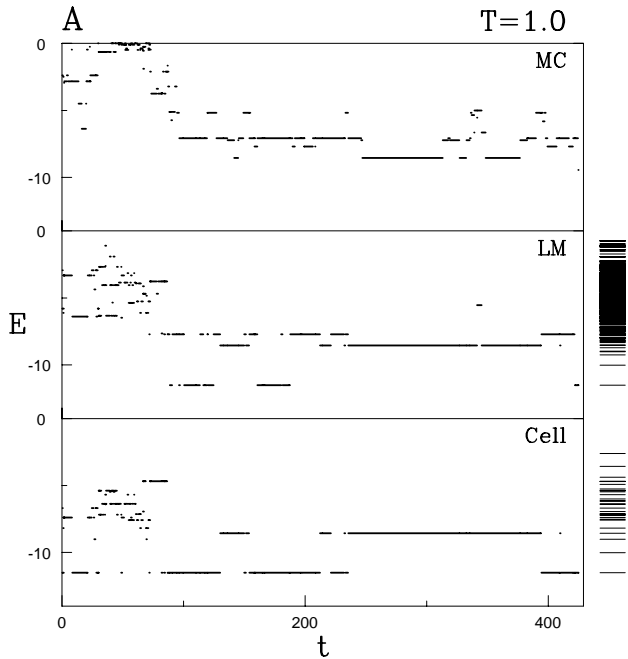}}
\caption{
Top: Energy  of states versus  number of the Monte Carlo steps in a
424-step long segment of one trajectory for sequence A.  Middle: Energy of
the local energy minima versus the Monte Carlo time.  The minima were
obtained by the steepest descent quenching from the states at the top.  The
energy spectrum of the minima is shown on the right.  Bottom: Energy of the
maximally compact states, obtained by mapping the LM's from the middle
panel, versus the Monte Carlo time. The energy spectrum of the cells is
shown on the right.}
\end{figure}

For sequence B,  local energy minima on the trajectories count overall less
than for A but, like for A, the proportions in equilibrium remain similar
to those found in the folding stage. A trap state for sequence B, however,
has a more substantial representation in equilibrium: it  accounts for 9\%
on trajectories and 17\% after the steepest descent quenching.   This trap
state has been discussed in reference\cite{8}.  Here, it is enough to
mention that  going from the trap state of sequence B  to the native state
requires full unfolding so this state forms a valley which is competing
with the native valley.  For sequence A, on the other hand,  the most
important trap is in the native valley and reaching the native state from
this trap requires only  partial unfolding.

\begin{figure}
\epsfxsize=3.4in
\centerline{\epsffile{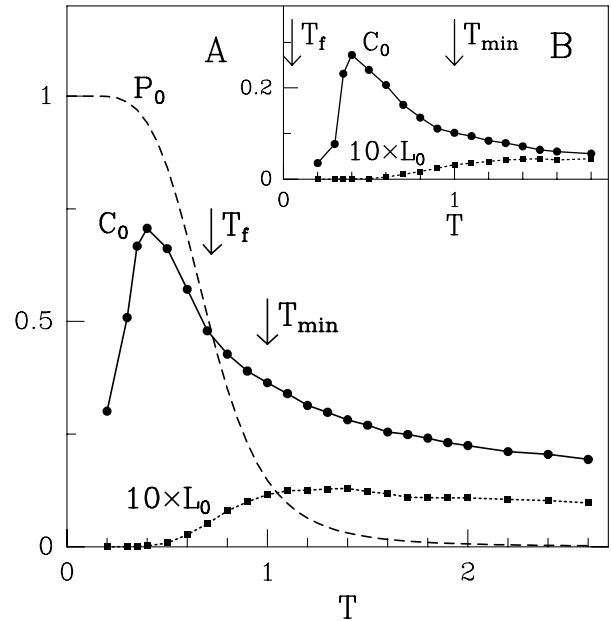}}
\caption{
Probability to find the native state before any mapping, $P_0$, after
one-stage mapping, $L_0$, and two-stage mapping $C_0$, as explained in the
text.  The main figure is for sequence A and the inset for B.  The values
of $T_{min}$ and $T_f$ are indicated.}
\end{figure}

\begin{figure}
\epsfxsize=3.4in
\centerline{\epsffile{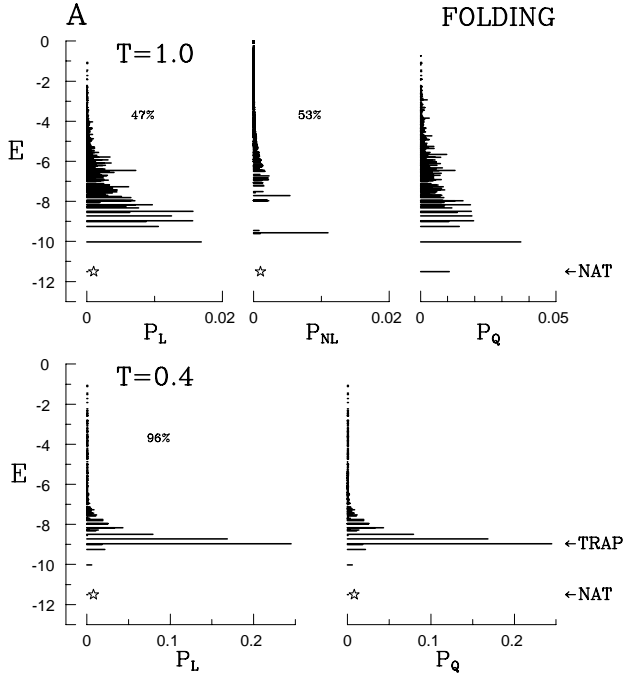}}
\caption{
Occupancy histograms for A based on 500 Monte Carlo trajectories that were
terminated at folding.  The asterisk marks the energy of the native state.
The top panels are for $T$=1 and the bottom panels for $T$=0.4.  $P_L$ and
$P_{NL}$ denote probabilities to find respectively LM's  and states which
are not LM's  in the Monte Carlo trajectories.  The numbers shown indicate
integrated probabilities.  $P_Q$  denotes probability to find an LM {\em
after} the steepest descent quenching.  The trap state accounts for about
1/4  of  the total time both before and after mapping to LM's. For sequence
B, the integrated occupancies at $T$=1 are 39\% and 61\% for $P_L$ and
$P_{NL}$ respectively. At $T$=0.4 the integrated $P_L$ is 77\%.}
\end{figure}
\vspace{10pt}

\noindent
{\bf Cell dynamics}

We now proceed to the various  ways to represent dynamics in the coarse
grained sense. We begin by discussing the description which is the most
economical  in presentation and the one that  has been  proposed in
references\cite{6,7}.  This is the  description based on the two-stage
mapping to cells, i.e. corresponding to the bottom of Figure 1.  When the
system leaves one cell and  arrives at another a connection between the two
cells is established. We count such connectivities in 500 trajectories
which terminate at folding and average to get connectivities.   The
connectivities can be represented in a matrix form.   The matrix is, in
principle, $31\times 31$ and it is symmetric. Most of the connectivities
are  weak or absent and a reduced matrix, by adopting a cutoff,  describes
the dynamics adequately. This is shown in Figure 4 which compares  the
dynamical matrix at $T_{min}$ and at a low temperature.  It is seen that
the good folder at the most favorite folding conditions generates a matrix
which involves many  direct connections to the native cell. The folding
funnel, in this description, consists of  the cells which are connected to
the native cell.  At low temperatures, and also for  the bad folder B at
any temperature,  the matrix looks more like the bottom panel of Figure 4:
there are much fewer direct connections of the low energy cells to the
native cell,  some connections become multiple step, or all connections
correspond to higher energy  motions that are not connected to the native
cell  ( this last mode, however, is not seen in Figure 4 since the system
is too small).  Finally, it should be pointed out that the cell dynamics
does not differentiates between the native cell and the trap state, because
the cell which is the closest to the trap happens to actually coincide with
the native cell for both sequences studied here.

\begin{figure}
\epsfxsize=3.2in
\centerline{\epsffile{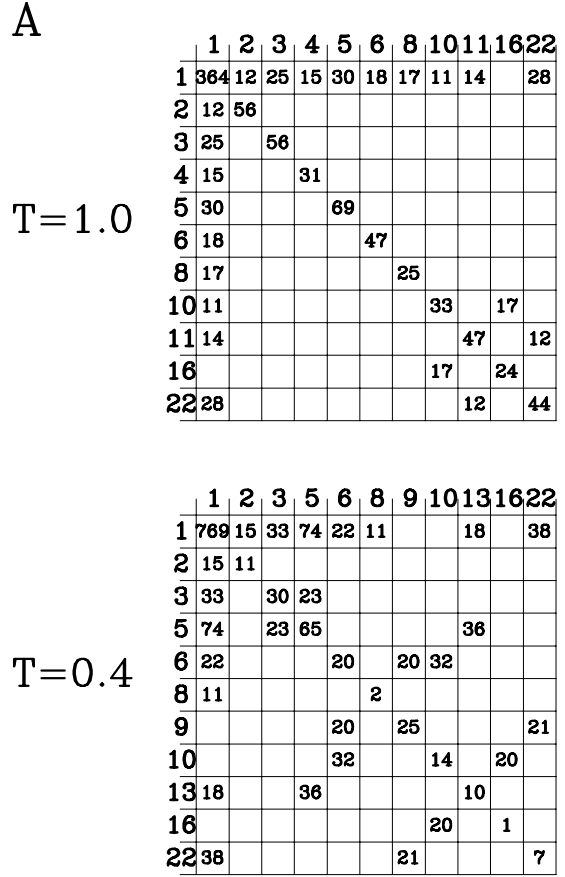}}
\caption{
Cell-to-cell connectivities for sequence A and temperatures indicated.  The
cell  labels correspond to the energy-wise rank ordering. Cell number 1 is
thus native.  The connectivities are normalized to 1000 and only those
larger than 10 are shown.  The diagonal terms indicate similarly normalized
values of the cell occupancies.}
\end{figure}

\noindent
{\bf Dynamics based on the local energy minima}

Consider now the one-stage mapping.  The connectivities are now determined
between the LM's as shown in the middle panel of Figure 1.  A matrix
representation of the results becomes impractical since too many states
are involved.  Instead we opted for the graphical representation as shown
in Figures 5 and 6 for sequences A and B respectively.  In these figures,
the vertical axis corresponds to energy. The positions of LM's in the
horizontal direction are chosen according to two criteria: 1) the
lines connecting them overlap as little as possible, 2) the states
corresponding to different clusters are shown separately (the cluster
analysis depends on the connectivity cutoff).
The x-axis coordinate is thus the conformation number, $N_c$, of the
local minimum. The labelling of the minima is well defined
and it is based on a computer generated listing.
The graphical horizontal placement of the LM on the figure,
however, is subjective and it is arranged in a way that
demonstrates the divisions into clusters of connectivities.
This subjectivness is due to the fact that we adopt a 2D
projection. In a many-dimensional space of the conformation
labels, the connectivity lines have an objective meaning.
The thickness of lines connecting the LM's is proportional to the frequency
of the appearance of the connection and the symbols corresponding to the
LM's have sizes controlled by probability to find these states.

\begin{figure}
\epsfxsize=3.4in
\centerline{\epsffile{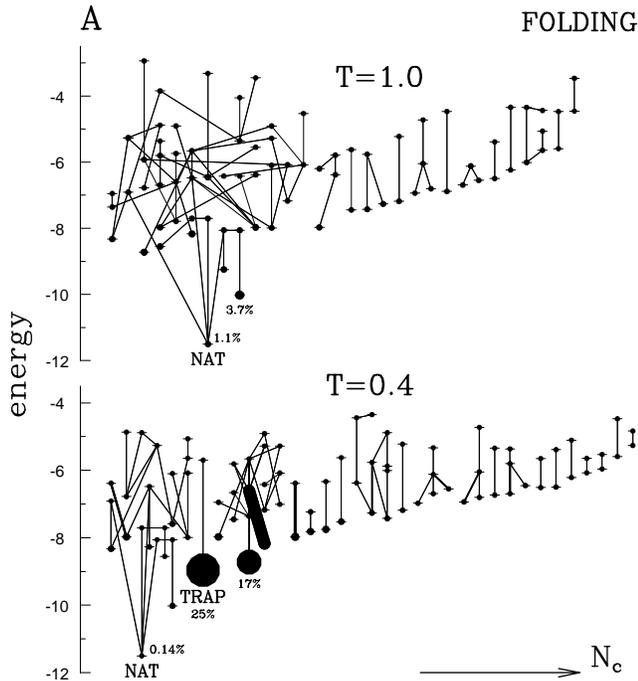}}
\caption{
Connectivities between the local energy minima for sequence A during
folding.  The connectivities are normalized to 1 and only those exceeding
0.001 are shown -- a full description would involve 495 LM's.  For $T$=1
and 0.4  the  connectivities displayed add up to 24\% and 73\%  of all
connectivities. The thickness of the connecting line indicates the
magnitude of  connectivity.  The size of the circle that locates an LM
indicates  its occupancy after quenching -- an analog of the diagonal
element in the matrix in Figure 4. The small print numbers shown indicate
the occupancy of the corresponding LM.  NAT indicates the native LM and
TRAP indicates the low temperature kinetic trap.}
\end{figure}

For the good folder, a well developed knot of states connected to the
native state is seen, in Figure 5, both at low temperatures and at those
which are good for folding.  We interpret this knot as the folding funnel.
In addition, other knots of relevant inter-valley motions are also present
-- the dynamics is indeed dominated by the funnel but it is not restricted
to it. At $T$=1, the trap state does not contribute to the effective
dynamics. It does contribute at $T$=0.4, however, but -- within the cutoff
adopted here - it is not connected to the funnel.

The graph of connectivities for the bad folder, shown in Figure 6,
indicates a much smaller knot connected to the native LM and no connections
to the native LM at the lower temperature.  All knots that are present are
at elevated energy states.  Thus this graphic representation clearly
distinguishes between good and bad folders.

Figure 7 is again for the good folder. It shows the graph of connectivities
at conditions of equilibrium, i.e. well past folding.  Again, the native LM
plays the dominant role. Furthermore, at low $T$'s the dynamics consists
primarily of transitions between the native state and a nearby LM which is
not the state which provides the most of kinetic trapping during folding at
low $T$'s.  

\begin{figure}
\epsfxsize=3.4in
\centerline{\epsffile{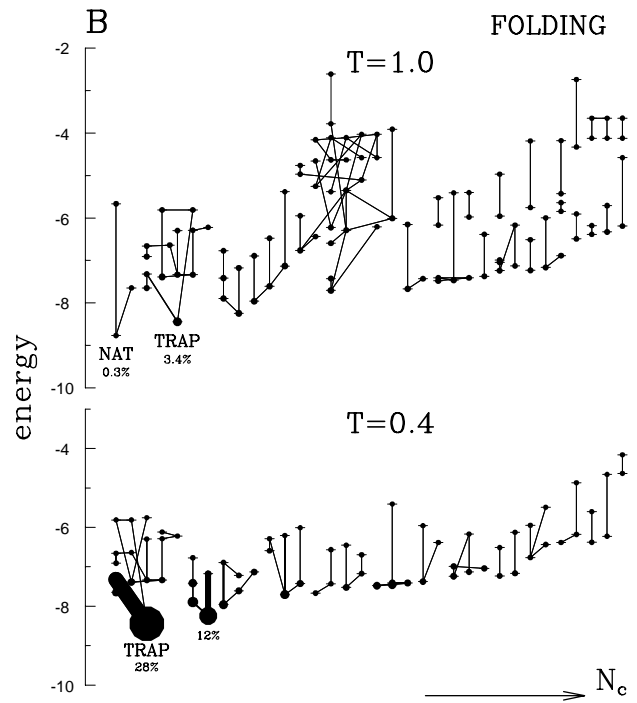}}
\caption{
Similar to Figure 5 but for sequence B.  Respectively,  24\% and 85\%
of all connectivities for
$T$=1 and 0.4 are displayed.}
\end{figure}

\begin{figure}
\epsfxsize=3.4in
\centerline{\epsffile{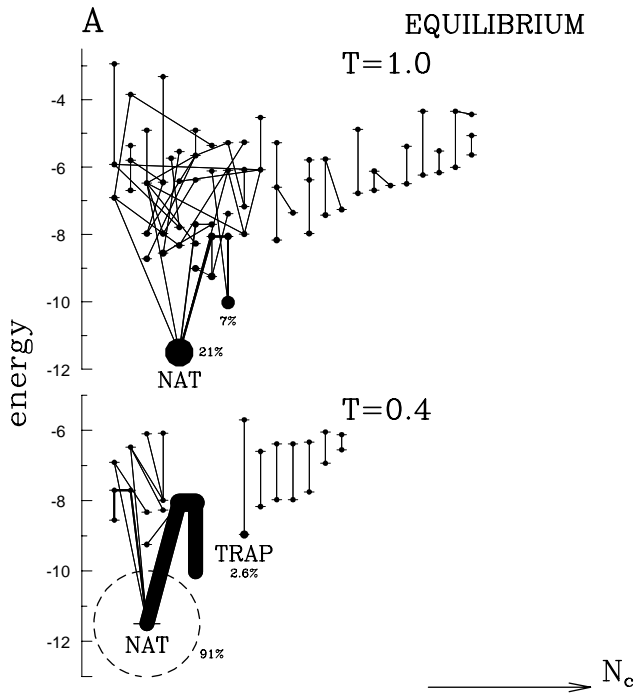}}
\caption{
Similar to Figure 5, and also for sequence A, but in equilibrium -- when
the trajectories are continued after folding (31\% of connections ata $T$=1
and 94\% at $T$=0.4).For sequence B, the graph in equilibrium is close to
the one corresponding to folding.}
\end{figure}

\begin{figure}
\epsfxsize=3.4in
\centerline{\epsffile{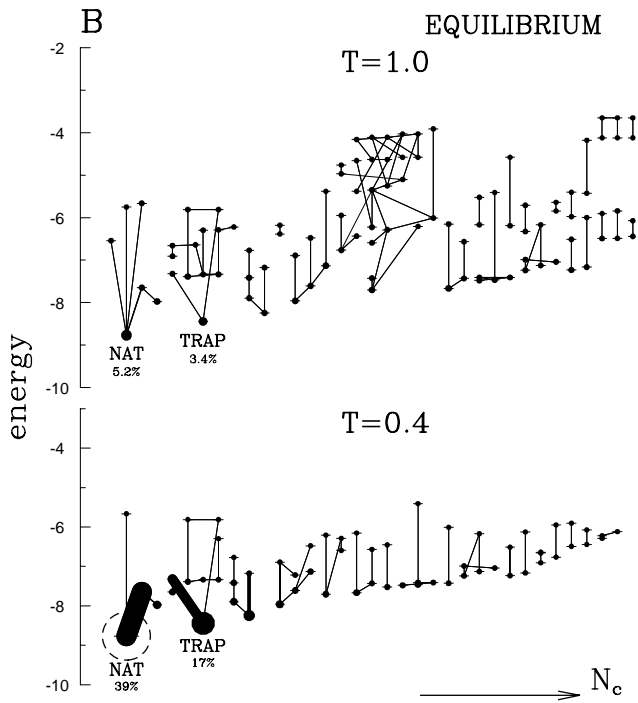}}
\caption{
Similar to Figure 7 but for sequence B.}
\end{figure}

Figure 8 shows the equilibrium graph of connectivities for
sequence B.  It clearly shows two, essentially equivalent but disconnected
knots one related to the native state and the other to the trap state. The
system then 'lives' essentially in these two valleys, which is consistent
with our understanding of the physics characterizing this sequence: for
sequence B, the native state can be reached from the trap state only
through a full unfolding.

\begin{figure}
\epsfxsize=3.4in
\centerline{\epsffile{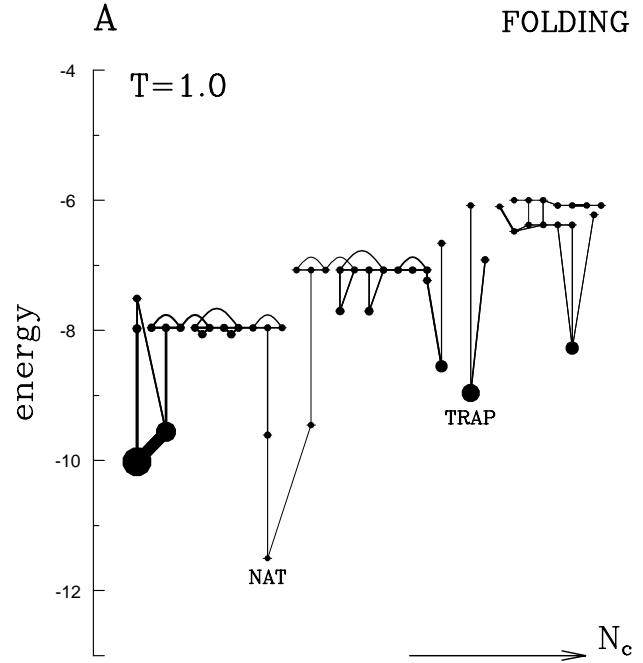}}
\caption{
Graph of connectivities between states on the Monte Carlo trajectory
for sequence A for the folding stage. The figure is based on
500 trajectories. There are many clusters and 3 of the most relevant ones
are displayed here. The native cluster consist of 34 states and the second
biggest cluster  of 15 states. 1000 of the most populated states, out of
15037, were monitored and the connectivity cutoff was 0.0002
of all monitored connectivities.}
\end{figure}

The overall look of the graphs  shown in Figures 5-8 is that of {\em
trees}, i.e. the graphs show very few loops.  This is not so when one does
not map the Monte Carlo states into local minima but just monitors
connectivities between the original states. In this case, many knots with
loops develop and an example of this is shown in Figure 9. This method of
monitoring the dynamics is not practical even for the 12-monomer system due
to the shear number of possible connections.

\section{{\bf  16-monomer sequences in two dimensions}}

We now come to more complex sequences.  Following reference\cite{6,7} we
consider 16-monomer sequences which have 802 075 conformations, of which 69
are cells. We focus on two sequences: R and DSKS'. The first is a good
folder, constructed by a rank-ordering procedure that assigns energies to
contacts, and the second is a bad folder which has been first studied by
Dinner et al.\cite{11}.  The values of the Gaussian contact energies have a
mean of about -1 in both cases.  The values of $T_f$ and $T_{min}$ are 1.15
and 1.2 for sequence A and 0.195 and 0.8  for DSKS'.  The plot of
$t_{fold}$ vs. $T$ for for sequence A is shown in Figure 10. One reason to
display it is that before no explicit care of the detailed balance
condition has been made (which affects the low $T$ branch of the curve
somewhat). More importantly, the figure shows that the low temperature data
agree with the Arrhenius law, $t_{fold} \sim exp(\delta E/T)$ with $\delta
E$ of about 3.7. 

The Arrhenius law has been obtained in the numerically exact studies of the
12-monomer sequences\cite{8}.  The barrier $\delta E$ in that case has been
associated with trajectories exiting from the most effective trap state and
ending in the native state. What determines $\delta E$ is the biggest
single step energy cost on the trajectory with the smallest overall
barrier. For the $N$=16 system A, we identify the trap state by studying
the biggest occupancies of the local energy minima at temperatures 0.4 and
0.3. The three most heavily represented traps (denoted as TRAP 1, TRAP 2,
and TRAP 3) are also shown in Figure 10, together with their corresponding
$\delta E$'s. All three are displayed because the Monte Carlo data yield
their occupancies to be of rather comparable values. The biggest $\delta
E$, of 3.6263, is associated with TRAP 1 which explains the value found by
fitting $t_{fold}$ to the Arrhenius law. The other two barriers have
similar but smaller values.

\begin{figure}
\epsfxsize=3.4in
\centerline{\epsffile{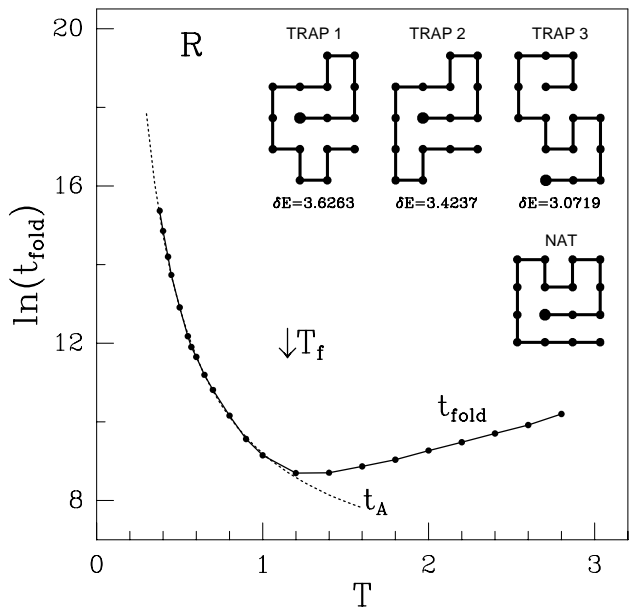}}
\caption{ Median folding time, based on 200 Monte Carlo trajectories, for
the 16-monomer sequence R -- the solid line. The dotted line corresponds to
the Arrhenius law with $\delta E$=3.7.  The conformations shown at the top
are the three most relevant kinetic trap states. The corresponding values
of $\delta E$ are written underneath.  The native conformation is denoted
by NAT. The first bead is shown enlarged.}
\end{figure}

We now consider coarse graining of the dynamics.  The results on the cell
dynamics have been reported before\cite{6,7} and here we focus only on the
dynamics based on the LM's.  Sequence R has 9103 LM's (out of which 2024
are U-shaped) and sequence DSKS' -- 9424 (2253 U-shaped). We generate 200
Monte Carlo trajectories that we map to LM's by the steepest descent
method. For each $T$ we determine which of these LM's belong to the top
1000 in terms of their occupational probability.  We then redo the runs and
monitor connectivities between the 1000 LM's. 

Figure 11 shows relevant portions of the connectivity graph for sequence R
at $T$=1.2 and 2.0 whereas Figure 12 is for $T$=0.8 and 0.6. The two
figures have been obtained by using a cutoff of 0.001 for a single
connectivity line with a normalization in which all monitored
connectivities add up to 1.  The tree which could be interpreted as the
folding funnel is most extended at $T$=1.2, i.e. for the most favorable
folding conditions. This tree sheds its branches on going both to high and
low temperatures. The transitions at low $T$ span much smaller energies
than at higher $T$'s. Furthermore, the low $T$ non-native trees are quite
elaborate. The looks of the low and high $T$ graphs are quite distinct then
and at $T_{min}$ the features of the low and high $T$ dynamics combine to
generate an involved funnel of states.

Figure 13 shows the corresponding graphs for DSKS' at $T$=1.2 and 0.6.
There is no tree of connections to the native state at any of these
temperatures. Instead there are many disconnected clusters that cover
small energy scales.

\begin{figure}
\epsfxsize=3.4in
\centerline{\epsffile{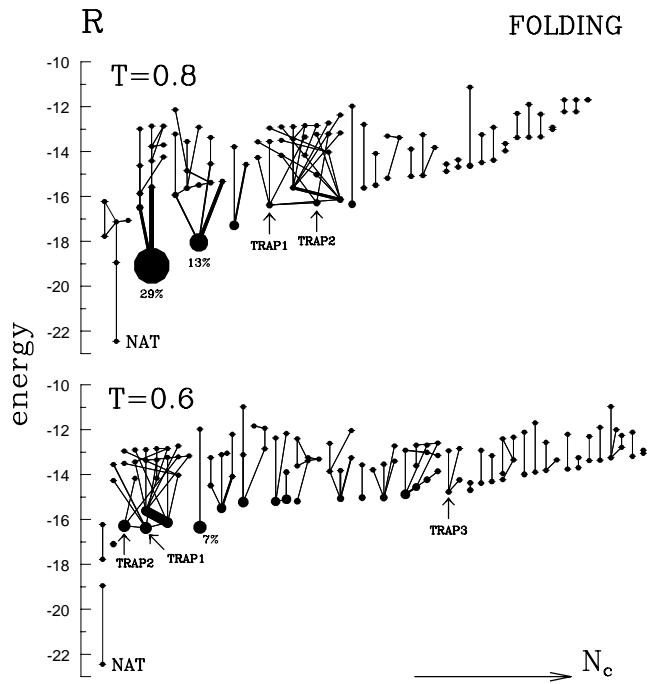}}
\caption{
Graph of connectivities for sequence R at $T$=0.8 and 0.6. The top
figure shows 58\% of all connectivities that were monitored and the
bottom figure -- 69\%. Other connectivity graphs are not shown.
The trap states are indicated.}
\end{figure}

\begin{figure}
\epsfxsize=3.2in
\centerline{\epsffile{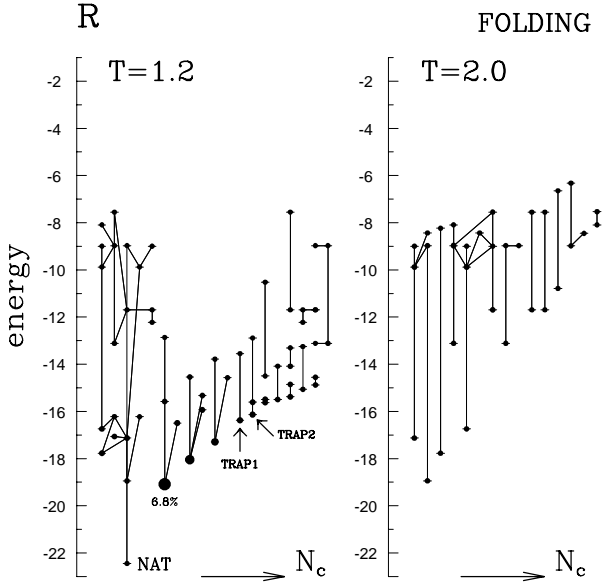}}
\caption{
Graph of connectivities for sequence R at $T$=1.2 and 2.
The first shows 19\% and the other 9\% of the connectivities.}
\end{figure}

\begin{figure}
\epsfxsize=3.2in
\centerline{\epsffile{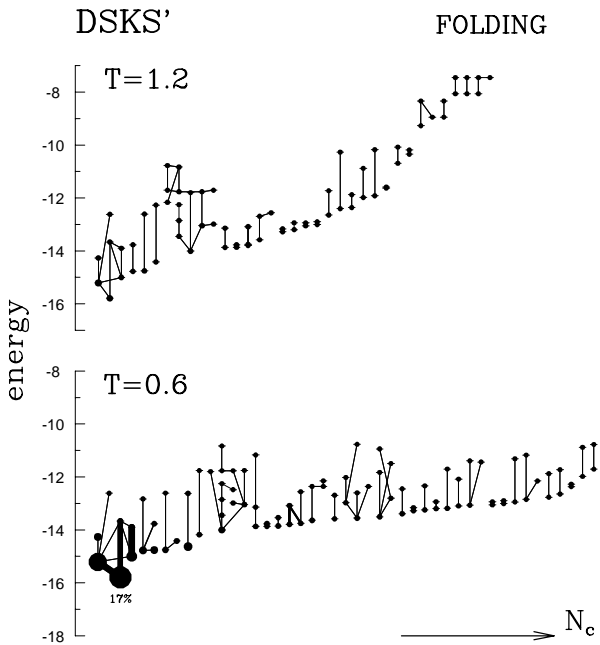}}
\caption{
Graph of connectivities for sequence DSKS'. For $T$=1.2 18\%
of the connectivities are shown, whereas for $T$=0.6 -- 71\%.}
\end{figure}

\section{{\bf 27-monomer sequence in three dimensions}}

The problems of the state monitoring compound when working with
heteropolymers in three dimensions. For a 27-monomer chain, such as
considered by Sali et al.\cite{12} and Shrivastava et al.\cite{13} one
cannot even enumerate all local energy minima, except for those which are
maximally compact, so we need a basis of states that relates only to the
states encountered.

We have constructed a sequence, C, by generating the 156 contact energies
from the Gaussian probability distribution with the mean of -2 and
assigning them to the target shown in ref.\cite{12}. The assignment has
been done as follows: the values of contact energies were rank ordered and
the strongest attracting couplings were allocated to the 28 contacts
present in the $3\times 3 \times 3$ target shape. The signs of the
remaining non-native contact energies were modified so that 50\% of them
were attractive and 50\% repulsive.  This way of the sequence design has
been demonstrated\cite{13} as leading to the fastest folding
characteristics. Our Monte Carlo based estimates for sequence C yield $T_f
\approx$ 2.57 and $T_{min} \approx$ 2.5.  At $T_{min}$, the median folding
time is very short, for 3$D$ sequences, -- of order 45000 steps which
minimizes the number of states to deal with.

\begin{figure}
\epsfxsize=3.4in
\centerline{\epsffile{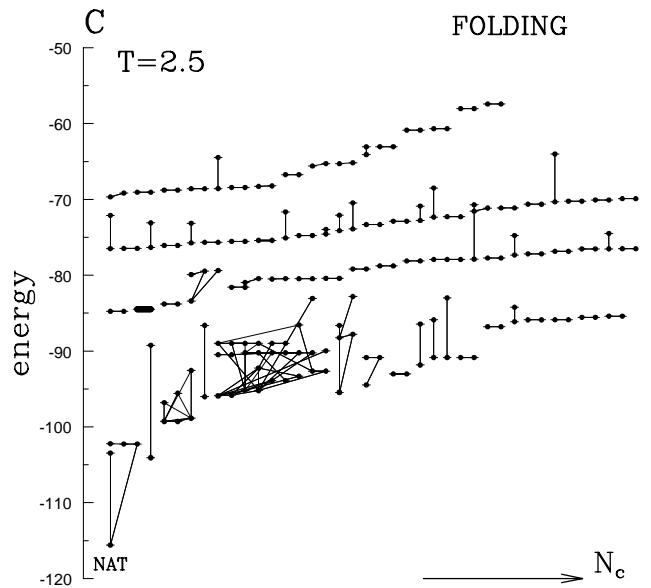}}
\caption{
Graph of connectivities for the three dimensional 27-monomer sequence C.
The cutoff of adopted here is 0.0005.  The basis of 1000 LM's used takes
into account about 23\% of the total Monte Carlo time.}
\end{figure}

In order to characterize the dynamics by the LM-based  connectivity graphs
we adopted the following procedure. First, we generated 100 folding
trajectories at $T_{min}$ and selected 30 of them which were the shortest.
The steepest-descent-based mapping was then applied to the selected
trajectories. For each of them, the number of LM's did not exceed 20 000 We
worked with a temporary basis of 20 000 local minima from which low
occupancy states were being removed during the process.  The end result was
to pick 1000 'finalists' -- the LM's which were populated the most.  The 30
runs were repeated again to determine the connectivities between the 1000
finalists. These are shown in Figure 14. The LM-based dynamics is seen to
be very fragmented with little structure which would be connected to the
native state. It is possible that delineating a native knot - due to the
enormous number of states - needs much more averaging over trajectories
than the number we could study.

Therefore, we considered another approach in which we do not monitor the
strengths of the connectivities but study the overlaps between the 30
trajectories, no matter how often a given link has appeared.  The resulting
graph of connectivities is shown in Figure 15. Here, we show only those
links which have appeared in at least 4 trajectories which represents the
dynamics in terms of 11 knots or clusters (there would be 45 clusters with
the cutoff of 2). The native knot is tree-like and it has a substantial
structure.  This suggests that when too many states are involved, an
overlap method of constructing the linkages may be preferable.

\begin{figure}
\epsfxsize=3.4in
\centerline{\epsffile{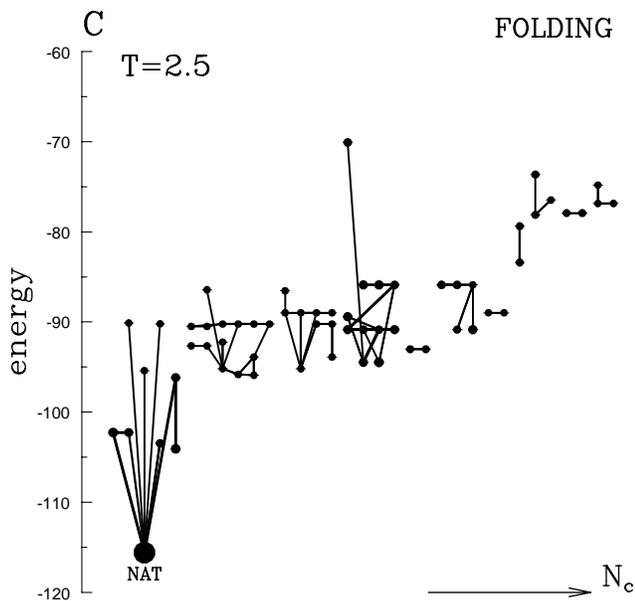}}
\caption{
Graph of connectivities for sequence C obtained by studying overlaps
between the trajectories.}
\end{figure}

We conclude that the steepest-descent based dynamics does allow to
distinguish between the good and bad folders. It provides a fairly
detailed  and meaningful representation of the dynamics, especially
in the case of small two-dimensional systems. 
The connectivity patterns and the emergence of structures that can be
identified with the folding funnel in good folders depend on the
Hamiltonian, the adopted dynamics, and on the kind of the lattice used.
Methods for three-dimensional heteropolymers need to be developed further.
It is expected that the procedures proposed in our paper will be even more
usefull when applied to off-lattice 3D models.
An alternative to the steepest descent based projection is to develop
coarse-graining methods which are not based on the mapping to the local
energy minima but instead, do statistical analysis of features in the
actual trajectories.
For instance, in a recent publication \cite{14}, Du et. al.  have proposed
to measure kinetic distances between conformations of heteropolymers in
terms of a `transition coordinate' which is related to
the probability to fold from a conformation without a preceding
unfolding.

It is a pleasure to acknowledge many stimulating discussions with J. R.
Banavar. Discussions with M. S. Li are also appreciated.
This work was supported by  KBN (Grant No. 2P03B-025-13).

\end{document}